\documentclass[twocolumn,conference]{IEEEtran}
\usepackage[T1]{fontenc}
\usepackage[latin9]{inputenc}
\usepackage{graphicx,float}
\usepackage{amsmath}
\usepackage{amsthm}
\usepackage{amssymb}
\usepackage{algorithm}
\usepackage{verbatim}
\usepackage{comment}
\usepackage{algpseudocode}
\usepackage[unicode=true,
 bookmarks=true,bookmarksnumbered=true,bookmarksopen=true,bookmarksopenlevel=1,
 breaklinks=false,pdfborder={0 0 0},pdfborderstyle={},backref=false,colorlinks=false]
 {hyperref}
\hypersetup{pdftitle={Your Title},
 pdfauthor={Your Name},
 pdfpagelayout=OneColumn, pdfnewwindow=true, pdfstartview=XYZ, plainpages=false}
\usepackage{stmaryrd,cuted}
\usepackage{multirow}

\makeatletter

\floatstyle{ruled}
\newfloat{algorithm}{tbp}{loa}
\providecommand{\algorithmname}{Algorithm}
\floatname{algorithm}{\protect\algorithmname}

\theoremstyle{plain}
\newtheorem{thm}{\protect\theoremname}
\theoremstyle{definition}
\newtheorem{defn}[thm]{\protect\definitionname}
\theoremstyle{plain}
\newtheorem{lem}[thm]{\protect\Lemmaname}
\theoremstyle{plain}
\newtheorem{example}[thm]{\protect\Examplename}

\usepackage[caption=false,font=footnotesize]{subfig}
\makeatother

\providecommand{\definitionname}{Definition}
\providecommand{\Lemmaname}{Lemma}
\providecommand{\theoremname}{Theorem}
\providecommand{\Examplename}{Example}

\begin{document}
\title{Power Allocation for Multi-Access Channel with Generalized Power Constraint} 

\author{Prashant Narayanan and  Lakshmi Narasimhan Theagarajan\\
Indian Institute of Technology Madras, Chennai 600036, India}
\maketitle
\begin{abstract}
We study the problem of decentralized power allocation in a multi-access channel (MAC) with non-cooperative users, additive noise of arbitrary distribution and a generalized power constraint, i.e., the transmit power constraint is modeled by an upper bound on $\mathbb{E}[\phi(|S|)]$, where $S$ is the transmit signal and $\phi(.)$ is some non-negative, increasing and bounded function.
The generalized power constraint captures the notion of power for different wireless signals such as RF, optical, acoustic, etc.
We derive the optimal power allocation policy when there a large number of non-cooperative users in the MAC.
Further, we show that, once the number of users in the MAC crosses a finite threshold, the proposed power allocation policy of all users is optimal and remains invariant irrespective of the actual number of users. 
We derive the above results under the condition that the entropy power of the MAC, $e^{2h(S)+c}$, is strictly convex, where $h(S)$ is the maximum achievable entropy of the transmit signal and $c$ is a finite constant corresponding to the entropy of the additive noise.
\end{abstract}


\section{\label{sec:Introduction}Introduction}

The uplink of various multi-user communication systems is characterized by the multi-access channel (MAC), where the communication medium is shared by several users simultaneously.
Non-cooperative power allocation in MAC is the technique through which each user chooses an average power for transmission without sharing any information to other users.
This power allocation should be such that total rate of all users are maximized while the interference is minimized.
Power allocation techniques for various kinds of wireless physical medium such as RF, optical, acoustic, etc., are known in the literature 
In each medium, the notion of transmit power is different.
For example, in RF systems, the average transmit power is constrained by an upper bound on $\mathbb{E}[|S|^2]$, whereas in optical systems, the power constraint is an upper bound on $\mathbb{E}[|S|]$, where the transmit signal is denoted by $S$. Therefore, power allocation techniques, designed so far in literature, depend on the kind of channel model and additive noise statistics.

In this paper, we present a generalized framework for power allocation that accommodates several physical layer models. 
The proposed framework introduces a general notion of power constraint as an upper bound on $\mathbb{E}[\phi(|S|)]$.
Further, this generalized framework can accommodate any distribution with finite entropy for the additive noise.
In contrast, most power allocation techniques known in the literature are applicable only for Gaussian noise.
The only condition that is assumed by the proposed framework on the MAC is that its entropy power $e^{2h(S)+2h(W)}$ is strictly convex, where $h(S)$ is the maximum achievable entropy of the transmit signal and $h(W)$ is the entropy of the additive noise.

From the generalized power allocation framework, we show that, once the number of users in the MAC crosses a finite threshold, there exists an invariant optimal power allocation policy.
Under this condition, this power allocation policy does not change even when new users or added or dropped out of the system.
Further, each user can compute this optimal power allocation policy with only the knowledge of the channel statistics of the users in the system and without any user-cooperation.
Thus, when sufficiently large number of users are present in the system, the proposed power allocation technique provides a low-complexity and optimal method of maximizing the MAC sum-rate.

\section{\label{sec:System-model}System model}

We consider a multi-access wireless system with $N$ transmitters and a receiver. At the receiver, the baseband samples at time $n$ is given by 
\begin{equation}
Y[n] = \sum_{i=1}^N H_i[n]S_i[n] + W[n],
\label{eq:Channel_model_equation}
\end{equation}
where $S_i[n]$ is the signal transmitted by the $i$th transmitter, $H_i[n]$ is the channel gain between the $i$th transmitter and the receiver, $W[n]$ is the additive noise of arbitrary distribution at the receiver. 
Model assumptions are:
\begin{enumerate}

\item The magnitudes of the channel gains are discrete, that is, $|H_i[n]| \in \mathcal{H}$, where $\mathcal{H}$ is a discrete, finite, non-negative and ordered set; $\mathcal{H}=\{h^{(1)}, \cdots, h^{(K)}\}$ and $h^{(1)}=0$.
\item The absolute value of the fade-coefficients for each transmitter $|H_i[n]|$ is a stationary ergodic process, with the stationary probability of $h^{(k)}$ for the $i$th transmitter denoted by $\pi_i(h^{(k)})$ and the $K\times 1$ dimensional vector of stationary probabilities is denoted by $\boldsymbol{\pi}_{i}$.

\item The differential entropy of the additive noise $W$ is finite, i.e., $h(W) < \infty$. 
\end{enumerate}

Each transmitter $i$ has the following power constraint
\begin{equation}\label{Time_avg_constraint}
\limsup_{T \rightarrow \infty}\frac{1}{T}\sum_{n = 1}^{T} \phi(|S_i[n]|) \leq \overline{G}_i,
\end{equation}
where the function $\phi: \mathbb{R} \rightarrow [0,\infty)$ is a non-negative, strictly increasing function with $\phi(0) = 0$ and $0<g_{min}\leq\overline{G}_i\leq g_{max}<\infty$ for all $i$. 
The function $\phi(.)$ models the {\em generalized power constraint}.
The time average sum rate of the system is given by,
\begin{equation}\label{Time_avg_Sum_rate}
\liminf_{T \rightarrow \infty}\frac{1}{T}\sum_{n = 1}^{T} \sum_{i=1}^{N} R_i[n],
\end{equation}
where $R_i[n]$ is the rate achieved by the $i$th transmitter. 

The goal of this work to provide efficient non-cooperative power allocation policies for each transmitter such that the average sum rate is maximized. That is, we design an efficient method where the $i$th transmitter chooses an amount of power given by $g_{i}(|H_{i}[n]|)$, for a given channel gain $H_{i}[n]$, such that the average sum rate is maximized without any cooperation among the users.
Once the optimal value of $g_{i}(H_i[n])$ is identified, the corresponding capacity achieving transmit signal distribution is employed by the $i$th transmitter for transmission in the $n$th time slot.

\section{\label{sec:Problem_statement}Problem formulation}

In this section, we use entropy power inequality (EPI) to obtain a lower bound on the sum rate and used it to formulate a sum-rate maximization problem for power allocation.  

\subsection{Lower bounds via EPI}
Using EPI for the sum-rate of the \eqref{eq:Channel_model_equation}, we get 
{
\begin{equation}
R[n] =\sum_{i=1}^N R_{i}[n] \geq  \frac{1}{2}\ln\left(1+\sum_{i=1}^N\frac{e^{2h(X_{i}[n])}}{e^{2h(W)}}\right),
\label{cap}
\end{equation}
}
\noindent where $h(X_{i}[n])$ is the differential entropy of $X_{i}[n] \triangleq H_i[n]S_i[n]$ and $X_{i}[n]$ follows the entropy maximizing distribution. 
Hence, $\mathbb{E}\left[\phi\left(|S_{i}[n]|\right)\right]\leq g_{i}(|H_i[n]|)$. For ease of exposition, henceforth, we drop the time index $n$.
The distribution of $X_{i}$ that maximizes the entropy, for a given $H_i$, is
{\small
\begin{equation}
  F_{X_i} = \underset{\substack{\{F_{X_i}(\cdot) ~:~ \mathbb{E}\left[\phi\left(|X_{i}/H_i|\right)\right] \leq g_{i}(|H_i|) \}}}{\arg\max} h(X_i),
  \label{Entropy_max}
 \end{equation}
}
where $F_{X_i}$ denotes all possible cumulative distribution functions (cdf) of $X_i$. Thus, from (\ref{cap}), we have
{\small
\begin{equation}\label{Applying_EPI}
\sum_{i=1}^{N} R_i[n] \geq\frac{1}{2}\ln\left(1+\sum_{i=1}^N\frac{e^{2h(X_{i})}}{e^{2h(W)}}\right).
\end{equation}
}
Let the scaled entropy power be denoted by 
\begin{equation}
N(|H_i|,g_i(|H_i|)) = \frac{e^{2h(X_{i})}}{e^{2h(W)}},
\end{equation}
where $N(h,g)$ is a function of the channel gain $h \geq 0$ and the average  power constraint $g \geq 0$. 

\begin{example}\label{Example}
$\phi(s) = |s|^p$ for $p>0$. 

In this case, the entropy maximizing distribution for (\ref{Entropy_max}) is the generalized Gaussian distribution whose probability density function (pdf) \cite{Dytso} is $f_{X_i}(x) = \frac{c_p}{\alpha}\exp\left(-\frac{|x|^p}{2\alpha^p}\right)$,
where $c_p = \frac{p}{2^{\frac{p+1}{p}}\Gamma\left(\frac{1}{p}\right)}$ and $\Gamma(\cdot)$ is the Gamma function. The maximum entropy is $h(X) = \frac{1}{p}\log\left(\frac{pe}{2c_p^p}\mathbb{E}[|X|^p]\right)$ and the scaled entropy function is $N(h,g) = \frac{ \left(\frac{pe}{2c_p^p}hg\right)^{\frac{2}{p}}}{e^{2H(W)}}$.

When $p=1$, this model represents an average amplitude constrained system such as the optical wireless communication system 
When $p=2$, this model represents an average power constrained system such as the RF wireless communication system.

\end{example} 

Thus, the proposed setup models a wide variety of communication systems and serves as a generalized model for power allocation in any multi-access channel.

\subsection{Power allocation policy}
A power allocation policy for the $i$th transmitter is denoted by an ordered set $\mathbf{F}_i$ consisting of $K$ cdf's, where the $k$th cdf is the conditional distribution of using power $g$ for channel gain $h^{(k)}$, i.e. $F_{i,k}(g) = \Pr(G_i \leq g | H_i = h^{(k)})$, where $G_i$ and $H_i$ are random variables that denote the transmit power and channel gain of the $i$th transmitter in a time slot, respectively. The set of all possible distributions $F_{i,k}$ on the set $[0,g_{max}]$ is denoted by $\mathcal{F}$.
The joint CDF of $G_i$ and $H_i$ is denoted by $Z_i(H_i, P_i) \triangleq\Pr(H_i = h^{(k)},G_i \leq g)= F_{i,k}(g)\pi_i(h^{(k)})$.  
Define the ordered set of functions, $\mathbf{Z}_i = \{Z_i(h^{(1)},\cdot), \cdots,Z_i(h^{(|\mathcal{H}|)},\cdot)\}$. 
Let $\mathbf{h} \triangleq [h_1, h_2, \cdots, h_N]^T$, $\mathbf{g} \triangleq [g_1, g_2, \cdots, g_N]^T$ and an ordered set $\mathbf{Z} \triangleq \{\mathbf{Z}_1, \mathbf{Z}_2, \cdots, \mathbf{Z}_N\}$. 
Using ergodicity, the average amount of power utilized in employing the policy $\mathbf{F}_i$ is 
\begin{equation}
\overline{{G}_{i}(\mathbf{Z}_i)}  = \mathbb{E}(G_i) = \sum_{k= 1}^{K}\int_{0}^{g_{max}}g\pi_i(h^{(k)}) dF_{i,k}(g).
\label{eq:Power_Feasibility}
\end{equation}
We define the objective function to be the lower bound on the average sum-rate as
$T(\mathbf{Z}) \triangleq \sum_{\mathbf{h}\in\mathcal{H}^N}\int_{\mathbf{g}}R(\mathbf{h},\mathbf{g})\prod_{i=1}^{N}dZ_{i}(h_{i},g_{i}),\label{eq:Average_rwrd_eqn}$
where, using \eqref{cap}, the sum-rate is defined as
\begin{equation}
R(\mathbf{h},\mathbf{g}) = \frac{1}{2}\ln\left(1+\sum_{i=1}^N N(h_i,g_i)\right)\label{eq:Capacity_eqn}.
\end{equation}
The set of all feasible policies of $i$th transmitter is defined as,
\begin{equation}
\mathcal{{Z}}_{i} \triangleq  \Big\{ \mathbf{Z}_{i}: 
\overline{{G}_{i}(\mathbf{Z}_{i})}\leq\overline{{G}}_{i}\Big\}.
\label{eq:Feasible_set}
\end{equation}
Let $\mathcal{{Z}}$ be the Cartesian product of $\mathcal{{Z}}_i$'s,
$\mathcal{{Z}} \triangleq \prod_{i=1}^N\mathcal{Z}_i$  and $\mathbf{Z}\in\mathcal{Z}$.
Each transmitter employs a feasible policy ($\mathbf{Z}_i^*\in\mathcal{Z}_i$) to maximize the average sum rate. 
Therefore, the problem of finding an efficient power allocation policy ($\mathbf{Z}^{*}$) is given by
\begin{equation}
\mathbf{Z}^{*} ~= ~~\underset{\mathbf{Z}\in\mathcal{Z}}{\arg\max} \quad T(\mathbf{Z}).
\label{eq:Problem_Formulation}
\end{equation}

In this work we show that under the assumption that the entropy functions $N(h^{(k)},g)$ is strictly convex for $k \geq 2$, there exist a policy $\mathbf{Z}^*$ which are the global maximizers of $T(\mathbf{Z})$ for all but finite number of users. In example (\ref{Example}), when $p<2$, the functions $N(hk^{()},g)$ is strictly convex for $k \geq 2$. We now provide the conditions under which this phenomenon occurs. Next we define the notion of Invariant policies.

\begin{defn}\label{Defn:Invariant policies}[\textbf{Invariant policy}]
A policy $\mathbf{Z}_i^* \in \mathcal{Z}_i$ is called Invariant policy of transmitter $i$, if there exists a integer $N^* \geq 1$, such that for each  $N\geq N^*$, the policies $\mathbf{Z}^*= \{\mathbf{Z}^*_1,\mathbf{Z}^*_2,\cdots,\mathbf{Z}^*_N\}$ satisfies (\ref{eq:Problem_Formulation}). Further the policies $\mathbf{Z}^*$ are called Invariant policies.
\end{defn}

\begin{defn}\label{Defn:Reg_cndtns}[{\em Regularity Conditions}]
(1) The probability of having strictly positive channel for any user is nonzero, i.e. 
$\pi(h^{(1)})<1-\eta$ for some $\eta\in(0,1)$, and (2) the entropy function $N(h,g)$ is strictly convex in $g$, for $h>0$ and $g \geq 0$.
\end{defn} 

\begin{thm}
\label{thm:IP_existence}
 When the regularity conditions in Definition \ref{Defn:Reg_cndtns} hold true, there exists a invariant power allocation policy whose pdf is given by
{\scriptsize
\begin{equation}
f^*_{i,k}(g)=\begin{cases}
1 & \text{ when } G_i=g_{max}, k > \tau\\\\
 
\frac{\overline{G}_i - \sum\limits_{l > \tau}g_{max}\pi_i(h^{(l)})}{g_{max}\pi_i(h^{(\tau)})}   & \text{ when } G_i=g_{max}, k = \tau  \\\\
 
1 - f_{i,\tau}(g_{max}) & \text{ when } G_i=0, k = \tau  \\\\
 
1 & \text{ when } G_i=0, k < \tau \\\\ 
0 & \text{ otherwise},
\end{cases}
\label{Invariant_pa}
\end{equation}
}  
where $ $
{\scriptsize
\begin{equation}\label{Def:tau}
\tau =\underset{1\leq k\leq K}{\arg\max}\left\{\sum\limits_{h > h^{(k)}}g_{max}\pi_i(h^{(k)}) \leq \overline{G}_i < \sum\limits_{h \geq h^{(k)}}g_{max}\pi_i(h^{(k)})\right\}.
\end{equation}
}

\end{thm}
\begin{proof}
Refer Appendix \ref{sec:Proof_IP_EXISTENCE}.
\end{proof}

\section{Conclusions and Future Scope}

\bibliographystyle{IEEEtran}
\bibliography{Rawnet}

\appendix
\begin{appendices}

\section{\label{sec:Proof_IP_EXISTENCE}Proof of theorem \ref{thm:IP_existence}}
To prove the existence of invariant policies as defined in Definition \ref{Defn:Invariant policies}, we first define  a set called the Universal Best Response (UBR) set for each transmitter $i$. We shall show that any invariant policy belongs to this set. We establish that, if there exist invariant policies, then under the regularity conditions, it must be (\ref{Invariant_pa}).
We also show that these policies are global maximizers of (\ref{eq:Problem_Formulation}).
%

\subsection{Universal Best-Response (UBR) Sets}\label{Best_Response_sets}

In this section we introduce the concept of UBR sets. 
Let $\mathbf{Z}_{-i}\in\mathcal{Z}_{-i}$ represent the power allocation policies of all transmitters except the $i$th transmitter and, where $\mathcal{Z}_{-i}=\prod\limits_{j=1,~j\neq i}^N\mathcal{Z}_{j}$. 
Let $\left(\mathbf{Z}_{i},\mathbf{Z}_{-i}\right) \in \mathcal{Z}$ denote the case where the $i$th transmitter employs the policy $\mathbf{Z}_{i}$ and the remaining transmitters use the policies given by $\mathbf{Z}_{-i} \in \mathcal{Z}_{-i}$. Similarly, we define the notations $(h_i,\mathbf{h}_{-i})$, where $\mathbf{h}_{-i} \in \mathcal{H}^{N-1}$ and $(g_i,\mathbf{g}_{-i})$, where $\mathbf{g}_{-i} \in [0,g_{max}]^{N-1}$.  Now, for a given $\mathbf{Z}_{-i}$, the best response power allocation policy $\mathbf{Z}^*_{i}$ of transmitter $i$ is the policy that satisfies,
\begin{equation}
\mathbf{Z}^*_{i} \in \arg\max_{\mathbf{Z}_i \in \mathcal{Z}_i}T\left(\mathbf{Z}_i,\mathbf{Z}_{-i}\right).\label{eq:Best_response_eqn}
\end{equation}
For a fixed $\mathbf{Z}_{-i}$, we have

\begin{equation}
T(\mathbf{Z}_i,\mathbf{Z}_{-i}) \triangleq \sum_{h\in\mathcal{H}}\int_{\mathbf{g}}C(h,\mathbf{Z}_{-i})dZ_{i}(h_{i},g_{i}),\label{eq:Best_response_Obj_fnc}
\end{equation}
where 
{\scriptsize
\begin{eqnarray}
C(h, \mathbf{Z}_{-i}) = \hspace{-3mm}
\sum_{\mathbf{h}_{-i}\in\mathcal{H}^{N-1}}\int_{\mathbf{g}_{-i}}\hspace{-2mm}R\left(\left( h, \mathbf{h}_{-i}\right), \left(g, \mathbf{g}_{-i}\right)\right)\hspace{-1mm}\prod_{j=1,~ j\neq i}^{N}dZ_{j}(h_{j},g_{j}),\nonumber
\end{eqnarray}
\begin{eqnarray}
R\left(\left( h, \mathbf{h}_{-i}\right), \left(g, \mathbf{g}_{-i}\right)\right) = \frac{1}{2}\ln\left(1+N(h_i,g_i) +\sum_{j=1,j \neq i}^N N(h_j,g_j)\right).
\end{eqnarray}
}

Given a system of $N$ transmitters and power allocation policies $\mathbf{Z}_{-i}$, we define $\mathcal{{B}}_N(\mathbf{Z}_{-i})$ as the set of all best response policies of user-$i$, i.e., 
\begin{equation}
\mathcal{{B}}_N(\mathbf{Z}_{-i}) = \left\{ \mathbf{Z}_{i}^* ~\Big |~ \mathbf{Z}_{i}^* =\underset{\mathbf{Z}_i\in\mathcal{Z}_i}{\arg\max} ~T\left(\mathbf{Z}_{i},\mathbf{Z}_{-i}\right)\right\}.
\label{bdef}
\end{equation}

Let $\mathcal{D}_{\eta}$ be the set of all distributions with a atleast a $\eta > 0$ probability of strictly positive channel. We formalize this as 
\begin{eqnarray}\label{Set_of_all_stat_distbn}
    \mathcal{D}_{\eta} = \Bigl\{\textbf{d}\triangleq [d(h^{(1)}),\cdots,d(h^{(K)})]^T~\Big|~\textbf{1}^T{\bf d}=1;\qquad\qquad\quad\nonumber\\
	~~0 \leq d(h^{(1)}) < 1-\eta,~ 0\leq d(h^{(k)}) \leq 1, ~k=2,\cdots,K, \Bigr\}.\nonumber
\end{eqnarray}

We define the UBR set of transmitter-$i$ (i.e., under any number of transmitters $N$ and any arbitrary stationary distributions for other transmitters) as

 {\small
 \begin{equation}\label{eq:DEFN:Set_of_all_Best_responses}
 \mathcal{M}_{i} = \hspace{-0.5mm}\bigcup_{N=1}^{\infty}~\bigcup_{\boldsymbol{\pi}_{-i}\in\mathcal{D}_{\eta}^{N}}\bigcup_{\textbf{Z}_{-i}\in\mathcal{Z}_{-i}} \hspace{-2.5mm}\mathcal{{B}}_{N+1}(\mathbf{Z}_{-i}), ~\text{where } \mathcal{Z}_{-i} =\hspace{-3mm}\prod\limits_{j=1,j\neq i}^{N+1}\mathcal{Z}_{j}. 
 \end{equation}
 }
 In Appendix \ref{propb}, we provide properties of the UBR set $\mathcal{M}_i$ under the regularity conditions. Next we show that invariant policies if they exist, lie in the UBR set.
 \begin{thm}\label{InV_belongs_to_Uni}
 Let $\mathbf{Z}_i^*\in \mathcal{Z}_i$ be an invariant policy of transmitter $i$, then $\mathbf{Z}_i^*\in \mathcal{M}_i$.
 \end{thm}
\begin{proof}
As $\mathbf{Z}_i^*\in \mathcal{Z}_i$ is an invariant policy, there exists an integer $N^* \geq 1$, such that, for all $N \geq N^*$, $\mathbf{Z}^*_{i} \in \arg\max_{\mathbf{Z}_i \in \mathcal{Z}_i}T\left(\mathbf{Z}_i,\mathbf{Z}_{-i}^*\right).$ Thus, from (\ref{bdef}) and  (\ref{eq:DEFN:Set_of_all_Best_responses}),  $ \mathbf{Z}^*_{i} \in \mathcal{{B}}_{N+1}(\mathbf{Z}_{-i})$ and $\mathbf{Z}_i^*\in \mathcal{M}_i$.
  
\end{proof}

\subsection{$\mathcal{M}$-dominating policies\label{bepsilon dominating}}

In this section, we define policies called $\mathcal{M}$-dominating policies. 
A policy $\mathbf{Z}_{i}^*$ of transmitter-$i$ is a $\mathcal{M}_{-i}$-dominating policy if it is the best response when the other transmitters employ policies from their UBR set. That is, 

\vspace{-3mm}
{\small 
\begin{eqnarray}
\mathbf{Z}_i^* \in \underset{\forall\ \mathbf{Z}_{i}\in\mathcal{{Z}}_{i},\forall\ \mathbf{Z}_{-i}\in\mathcal{{M}}_{-i}}{\arg\max} T\left({Z}_{i},\mathbf{Z}_{-i}\right), \mathcal{M}_{-i} \triangleq \prod^N_{j=1, j\neq i}\mathcal{M}_{i}.\label{eq:B_Dominating_policy_eqn}
\end{eqnarray}
}
{\bf Definition}: A policy $\mathbf{Z}^* \in \mathcal{Z}$ is said to be $\mathcal{M}$-dominating policy when $\mathbf{Z}_i^*$ is $\mathcal{M}_{-i}$ dominating for all users $i$. 

\begin{thm}[\cite{vlc_paper}]
\label{thmDSE_are_Maxima}
The $\mathcal{M}$-dominating policy is the global maximizer of $\underset{\mathbf{Z}\in\mathcal{Z}}{\arg\max}~ T(\mathbf{Z})$.
\end{thm}

In Appendix \ref{Existence_proof}, we show that the policies defined in (\ref{Invariant_pa}) are $\mathcal{{M}}$-dominating policies when the number of users exceeds some fixed positive integer. This also establishes that that the policies (\ref{Invariant_pa}) are invariant policies.

\section{Invariant policies: Necessary conditions}
\label{Inv_form}
In this section, we prove that the invariant policies, if they exist satisfy (\ref{Invariant_pa}). Before showing this, we give the following preliminary definitions and results.
The following bounds are repeatedly used in this work.
{\scriptsize
\begin{equation}\label{log_inequailities}
x- \frac{x^2}{2} \leq \log(1 + x) \leq x, \; \forall x\geq 0. 
\end{equation} 
}

Let $\Omega:\mathcal{H}^N \times \mathcal{G}^N \rightarrow \mathbb{R}$ be any real valued function. Then given the policies $\mathbf{Z}$, we denote the Expectation of the function $\Xi$ as,
{\scriptsize
\begin{equation}\label{Defn:Expectation}
\mathbb{E}[\Omega] = \sum_{\mathbf{h}\in\mathcal{H}^{N}}\int_{\mathbf{g}}\Omega(\mathbf{h},\mathbf{g})\prod_{j=1}^{N}dZ_{j}(h_{j},g_{j})
\end{equation}
}
and the expectation with only the distributions $\mathbf{Z}_j,\; j \neq i$ for a fixed $h_i$ and $g_i$ of transmitter $i$ as,
{\scriptsize
\begin{equation}\label{Expectation_i}
\mathbb{E}_{-i}[\Omega(h_i,g_i)] = \sum_{\mathbf{h_{-i}}\in\mathcal{H}^{N-1}}\int_{\mathbf{g_{-i}}}\Omega(\mathbf{h},\mathbf{g})\prod_{j=1,j\neq i}^{N}dZ_{j}(h_{j},g_{j})
\end{equation}
}
\noindent{\bf Definition}: For a given $(\mathbf{Z}_i,\mathbf{Z}_{-i})$, we define an objective function for the $i$th user as
\begin{equation}
T_{i}\left(\mathbf{Z}_{i},\mathbf{Z}_{-i}\right) = \mathbb{E}[U_{i}(\mathbf{h},\mathbf{p})].\label{eq:Alternative_Expected_rwrd}
\end{equation}

where 
\begin{equation}
U_{i}(\mathbf{h},\mathbf{p}) = \frac{1}{2}\log\left(1+\frac{N(h_i,g_i)}{1+\sum_{j=1,j\neq i}^{N}N(h_j,g_j)}\right),\label{eq:Alternative_rwrd}
\end{equation}
Further, 
\begin{equation}
    U_i(\mathbf{h},\mathbf{p}) = R(\mathbf{h},\mathbf{p})-\frac{1}{2}\log\left(1+\sum_{j=1,j\neq i}^N N(h_j,p_j)\right).\nonumber 
\end{equation}
Hence, $\underset{\mathbf{Z}_i\in\mathcal{Z}_i}{\arg\max} ~T\left(\mathbf{Z}_{i},\mathbf{Z}_{-i}\right)=\underset{\mathbf{Z}_i\in\mathcal{Z}_i}{\arg\max} ~T_i\left(\mathbf{Z}_{i},\mathbf{Z}_{-i}\right)$.
Therefore, the set $\mathcal{{B}}_N(\mathbf{Z}_{-i})$ can also be expressed as
\begin{equation}
\mathcal{{B}}_N(\mathbf{Z}_{-i}) = \left\{ \mathbf{Z}_{i}^* ~\Big |~ \mathbf{Z}_{i}^* =\underset{\mathbf{Z}_i\in\mathcal{Z}_i}{\arg\max} ~T_i\left(\mathbf{Z}_{i},\mathbf{Z}_{-i}\right)\right\}.
\label{bdef1}
\end{equation}

Now we establish that any invariant policy must satisfy (\ref{Invariant_pa}). Let $\mathbf{Z}_i^*$ be an invariant policy for transmitter $i$. Then by definition (\ref{Defn:Invariant policies}), for some positive integer $N_0$ and any $N \geq N_0$, $\mathbf{Z}_i^* = \arg\max\limits_{\mathbf{Z}_i \in \mathcal{Z}_i}T(\mathbf{Z}_i,\mathbf{Z}_{-i}^*)$. Further using (\ref{bdef1}) and (\ref{eq:Alternative_Expected_rwrd}), we have $\mathbf{Z}_i^* = \arg\max\limits_{\mathbf{Z}_i \in \mathcal{Z}_i}T_i(\mathbf{Z}_i,\mathbf{Z}_{-i}^*) = \arg\max\limits_{\mathbf{Z}_i \in \mathcal{Z}_i}\mathbb{E}[U_i(\mathbf{h},\mathbf{p})]$. Now using (\ref{log_inequailities}) and (\ref{eq:Alternative_rwrd}) , we have that 

{\scriptsize
\begin{eqnarray}\label{Inv_1}\nonumber
&&\mathbb{E}\left[\frac{N(h_i,p_i)}{1 +\sum_{j=1,j\neq i}^N N(h_j,p_j)}\right]-\mathbb{E}\left[\left(\frac{N(h_i,p_i)}{1 +\sum_{j=1,j\neq i}^N N(h_j,p_j)}\right)^2\right]\\ 
&\leq & \mathbb{E}[U_i(\mathbf{h},\mathbf{p})] \leq \mathbb{E}\left[\frac{N(h_i,p_i)}{1 +\sum_{j=1,j\neq i}^N N(h_j,p_j)}\right].
\end{eqnarray}
}
From Theorem (\ref{InV_belongs_to_Uni}) and Theorem (\ref{lem:Chernoff_bds}), we have from (\ref{Inv_1}),
{\scriptsize
\begin{eqnarray}\label{Inv_2}\nonumber
&&\mathbb{E}\left[\frac{c_1 N(h_i,p_i)}{N}\right]-\mathbb{E}\left[c_2\left(\frac{N(h_i,p_i)}{N}\right)^2\right]
\leq  \mathbb{E}[U_i(\mathbf{h},\mathbf{p})] \\
&\leq & \mathbb{E}\left[\frac{c_1N(h_i,p_i)}{N}\right].
\end{eqnarray}
}
for some suitable strictly positive constants $c_1, c_2$. Thus from (\ref{Inv_2}), we get
\begin{equation}\label{Inv_3}
\lim_{N \rightarrow \infty}\arg\max\limits_{\mathbf{Z}_i \in \mathcal{Z}_i}N[U_i(\mathbf{h},\mathbf{p})] =\arg\max\limits_{\mathbf{Z}_i \in \mathcal{Z}_i}\mathbb{E}[N_i(h_i,g_i)]
\end{equation}
As $\mathbf{Z}_i^*$ is a invariant policy, by definition (\ref{Defn:Invariant policies}), we have from (\ref{Inv_3})
{\scriptsize
\begin{equation}\label{Inv_4}
\mathbf{Z}_i^* \in \arg\max\limits_{\mathbf{Z}_i \in \mathcal{Z}_i}\mathbb{E}[N_i(h_i,g_i)].
\end{equation}
}
Now using (\ref{Inv_4}), and strict convexity of $N(h,g)$ for $g \geq 0$ and all $h >0$, we show that $\mathbf{Z}_i^*$ satisfies (\ref{Invariant_pa}). Let 
{\scriptsize
\begin{equation}\label{Inv_lag_1}
L_1(\lambda,\mathbf{Z}_i) = \mathbb{E}[N_i(h_i,g_i)-\lambda g_i] + \lambda G_i
\end{equation}
}
If there exist a $\lambda \geq 0$ and some policy $\mathbf{Z}_i^*$ such that,
{\scriptsize
\begin{equation}
 \mathbf{Z}_i^* \in \arg\max\limits_{\substack{\{\mathbf{Z}_i ~:~ Z_i(h^{(k)},g_i) = \pi(h^{(k)})F_{i,k}(g)),\\\forall \;F_{i,k} \in \mathcal{F},\forall\;  1 \leq k \leq K \}}} L_1(\lambda, \mathbf{Z}_i),
  \label{Dual_opt_inv}
 \end{equation}
}
and $G_i(\mathbf{Z}_i^*)= \overline{G}_i$, then $\mathbf{Z}_i^*$ satisfies (\ref{Inv_4}). It can be easily shown that  (\ref{Dual_opt_inv}) holds if and only if $\forall \; k \geq 1$ 
{\scriptsize
\begin{equation}
 F_{i,k}^* \in \arg\max\limits_{\substack{\{F_{i,k} \in \mathcal{F} \}}} \int_{g}\left(N(h^{(k)},g)-\lambda g\right)dF_{i,k}(g),
  \label{Dual_opt_inv1}
 \end{equation}
}
By choosing $\mathbf{Z}_i^*$ as in (\ref{Invariant_pa}) and $\lambda= \frac{N(h^{(\tau)},g_{max})}{g_{max}}$, we have 
{
$\int\limits_{g=0}^{\infty}\left(N(h^{(k)},g)-\lambda g\right)dF^*_{i,k}(g)$
\begin{equation}\label{Dual_inv1}
= \begin{cases}
0 & k \leq \tau \\
N(h^{(k)},g_{max})-N(h^{(\tau)},g_{max})& k > \tau
\end{cases} 
\end{equation}
}
Further by the strict convexity of $N(h^{(k)},g)$ for all $g \geq 0$, for each $k > 1$, we have 
{\scriptsize
$N(h^{(k)},g)-\lambda g = $ $ \left(N(h^{(k)},g_{max}) - N(h^{(\tau)},g_{max})\right) \frac{g}{g_{max}}$
\begin{equation}\label{Dual_inv2}
 \leq \begin{cases}
 0 & k \leq \tau \\
N(h^{(k)},g_{max})-N(h^{(\tau)},g_{max})& k > \tau.
\end{cases}
\end{equation}  
}
Using (\ref{Dual_inv1}) and (\ref{Dual_inv2}), we can easily establish (\ref{Dual_opt_inv1}).  This establishes the result that if there exist a invariant policy $\mathbf{Z}_i^*$ for each transmitter $i$, it must be (\ref{Invariant_pa}). In the next section, we establish that (\ref{Invariant_pa}) are $\mathcal{M} - $dominating policies for each transmitter $i$ for almost all but finite number of users.   
\section{\label{Existence_proof}Proof of existence of $\mathcal{M}$-dominating policies}

\noindent{\bf Definition}: Given a power policy $\mathbf{Z}_i$, let $\mu_i \triangleq \mathbb{E}[N(h_i,g_i)]$ and $\mu \triangleq \inf_{i \geq 1}\mu_i$. In Appendix (\ref{propb}), we show that $\mu >0$ under the regularity conditions.
Here we will show here that there exist a $N^* \geq 1$, such that for all $N \geq N^*$ and $1 \leq i \leq N$, $\mathbf{Z}_i^* $ as defined in (\ref{Invariant_pa}) is $\mathcal{M}_{-i}$ dominating policy. To establish this we show that if Let $\mathbf{Z}_j \in \mathcal{M}_j$, for all $1\leq j \leq N$, then for all $N \geq N^*$, $\mathbf{Z}_i^* \in \arg\max_{\mathbf{Z}_i \in \mathcal{Z}_i}T_i(\mathbf{Z}_i,\mathbf{Z}_{-i})$. We do so by considering the Lagrangian
{\scriptsize
\begin{equation}\label{Lag_defn}
L(\lambda, \mathbf{Z}_i) = T_i(\mathbf{Z}_i,\mathbf{Z}_{-i}) - \lambda\left(G_i(\mathbf{Z}_i)- \overline{G}_i)\right)
\end{equation}
}
and the dual problem
{\scriptsize
\begin{equation}
  \hat{\mathbf{Z}}_i \in \arg\max\limits_{\substack{\{\mathbf{Z}_i ~:~ Z_i(h^{(k)},g_i) = \pi(h^{(k)})F_{i,k}(g)),\\\forall \;F_{i,k} \in \mathcal{F},\forall\;  1 \leq k \leq K \}}} L(\lambda, \mathbf{Z}_i),
  \label{Dual_opt}
 \end{equation}
}
 Further we define for each $h_i \in \mathcal{H}$, the function
{\scriptsize
\begin{equation}\label{Lag_breakup}
V(h_i,\lambda,g) = \mathbb{E}_{-i}\left[\log\left(1 +\frac{N(h_i,g)}{1 + \sum_{j \neq i} N(h_j,g_j)}\right)\right]-\lambda g,
\end{equation}
}
where
{\scriptsize
\begin{eqnarray}
&&\mathbb{E}_{-i}\left[\log\left(1 +\frac{N(h_i,g)}{1 + \sum_{j \neq i} N(h_j,g_j)}\right]\right)\\\nonumber
&=& \sum_{\mathbf{h_{-i}}\in\mathcal{H}^{N-1}}\int_{\mathbf{g}_{-i}}U_i(\mathbf{h},\mathbf{g})\prod_{j=1}^{N}dZ_{j}(h_{j},g_{j}).  
\end{eqnarray} 
}

We have a joint distribution $\hat{\mathbf{Z}}_i$ with $\hat{Z}_i(h^{(k)},g_i) =\pi_i(h^{(k)})\hat{F}_{i,k}(g_i)$ satisfying (\ref{Dual_opt}) if and only for all $h_i \in \mathcal{H}$, and all $F \in \mathcal{F}$
{\scriptsize
\begin{equation}\label{Dual_chk}
\int_{0}^{g_{max}}V(h_i,\lambda)d\hat{F}_{i,k}(g) \geq \int_{0}^{g_{max}}V(h_i,\lambda)dF(g).
\end{equation}  
}
We will establish that there exist a Lagrangian $\lambda > 0$ and a natural number $N^*$ such that $\mathbf{Z}_i^*$ as defined in (\ref{Invariant_pa}) satisfies the conditions (\ref{Dual_chk}) and thus the condition (\ref{Dual_opt}) for all $N \geq N^*$. As $\overline{G}_i(\mathbf{Z}_i^*) = \overline{G}_i$, we have from \cite{boyd2004convex}, that $\mathbf{Z}_i^* \in \arg\max_{\mathbf{Z}_i \in \mathcal{Z}_i}T_i(\mathbf{Z}_i,\mathbf{Z}_{-i})$ for $N \geq N^*$. 
Let
{\scriptsize
\begin{equation}\label{lambda_star}
\lambda =  \frac{1}{g_{max}}\mathbb{E}_{-i}\left[\log\left(1 +\frac{N(h^{(\tau)},g_{max})}{1 + \sum_{j \neq i} N(h_j,g_j)}\right)\right],
\end{equation}
}
then from (\ref{Dual_chk}) and using the policy $\mathbf{Z}_i^*$ as defined in (\ref{Invariant_pa}), we verify the following conditions
{\scriptsize
\begin{eqnarray}\label{Dual_condtn_1}
 0 &\geq& \int_{0}^{g_{max}}\mathbb{E}_{-i}\left[\log\left(1 +\frac{N(h^{(k)},g)}{1 + \sum_{j \neq i} N(h_j,g_j)}\right)\right]dF(g)\\\nonumber
&-&\int_{0}^{g_{max}}\frac{g}{g_{max}}\mathbb{E}_{-i}\left[\log\left(1 +\frac{N(h^{(\tau)},g_{max})}{1 + \sum_{j \neq i} N(h_j,g_j)}\right)\right]dF(g), \\\nonumber
&\forall& F(\cdot) \in \mathcal{F}, 1 \leq k \leq \tau
\end{eqnarray}
}
 and
{\scriptsize
\begin{eqnarray}\label{Dual_condtn_2}
&&\mathbb{E}_{-i}\left[\log\left(1 +\frac{N(h^{(k)},g_{max})}{1 + \sum_{j \neq i} N(h_j,g_j)}\right)\right] \\\nonumber
&-& \mathbb{E}_{-i}\left[\log\left(1 +\frac{N(h^{(\tau)},g_{max})}{1 + \sum_{j \neq i} N(h_j,g_j)}\right)\right]\\\nonumber
 &\geq& \int_{0}^{g_{max}}\mathbb{E}_{-i}\left[\log\left(1 +\frac{N(h^{(k)},g)}{1 + \sum_{j \neq i} N(h_j,g_j)}\right)\right]dF(g)\\\nonumber
&-&\int_{0}^{g_{max}}\frac{g}{g_{max}}\mathbb{E}_{-i}\left[\log\left(1 +\frac{N(h^{(\tau)},g_{max})}{1 + \sum_{j \neq i} N(h_j,g_j)}\right)\right]dF(g), \\\nonumber
&\forall& F(\cdot) \in \mathcal{F}, \tau + 1 \leq k \leq K.
\end{eqnarray}
}
In Lemma (\ref{Lem:Dual_condtn1}), we establish there exist a number $N_1$, such that for all $N \geq N_1$, 
{\scriptsize
\begin{eqnarray}\label{Dual_condtn_3}
 0 &\geq& \mathbb{E}_{-i}\left[\log\left(1 +\frac{N(h^{(k)},g)}{1 + \sum_{j \neq i} N(h_j,g_j)}\right)\right]\\\nonumber
&-&\frac{g}{g_{max}}\mathbb{E}_{-i}\left[\log\left(1 +\frac{N(h^{(\tau)},g_{max})}{1 + \sum_{j \neq i} N(h_j,g_j)}\right)\right], \\\nonumber
&\forall& g \in [0,g_{max}], 1 \leq k \leq \tau.
\end{eqnarray}
}
Integrating (\ref{Dual_condtn_3}) from $0$ to $g_{max}$ with respect to any distribution $F \in \mathcal{F}$, we get the desired result (\ref{Dual_condtn_1}) for all $N \geq N_1$. Similarly in Lemma (\ref{Lem:Dual_condtn2}), we establish there exist a number $N_2$, such that for all $N \geq N_2$,   
{\scriptsize
\begin{eqnarray}\label{Dual_condtn_4}
&&\mathbb{E}_{-i}\left[\log\left(1 +\frac{N(h^{(k)},g_{max})}{1 + \sum_{j \neq i} N(h_j,g_j)}\right)\right] \\\nonumber
&-& \mathbb{E}_{-i}\left[\log\left(1 +\frac{N(h^{(\tau)},g_{max})}{1 + \sum_{j \neq i} N(h_j,g_j)}\right)\right]\\\nonumber
 &\geq& \mathbb{E}_{-i}\left[\log\left(1 +\frac{N(h^{(k)},g)}{1 + \sum_{j \neq i} N(h_j,g_j)}\right)\right]\\\nonumber
&-&\frac{g}{g_{max}}\mathbb{E}_{-i}\left[\log\left(1 +\frac{N(h^{(\tau)},g_{max})}{1 + \sum_{j \neq i} N(h_j,g_j)}\right)\right], \\\nonumber
&\forall& g \in [0,g_{max}], \tau + 1 \leq k \leq K.
\end{eqnarray}
}
Integrating (\ref{Dual_condtn_4}) from $0$ to $g_{max}$ with respect to any distribution $F \in \mathcal{F}$, we get the desired result (\ref{Dual_condtn_2}) for all $N \geq N_2$. Thus by choosing $N^* = \max \{N_1, N_2\}$, we have both conditions (\ref{Dual_condtn_1}) and (\ref{Dual_condtn_2}) hold  forall  $N \geq N^*$. Thus for all $N \geq N^*$, the policy $\mathbf{Z}_i^*$ defined in (\ref{Invariant_pa}) satisfies (\ref{Dual_chk}) and thus (\ref{Dual_opt}). As the policy $\mathbf{Z}_i^*$ also satisfies $\overline{G}_i(\mathbf{Z}_i^*) = \overline{G}_i$ and $\lambda > 0$, we have that $\mathbf{Z}_i^* = \arg\max_{\mathbf{Z}_i \in \mathcal{Z}_i}T_i(\mathbf{Z}_i,\mathbf{Z}_{-i})$ for all $N \geq N^*$. Further as the policies of other transmitters $j  \neq i$ satisfy $\mathbf{Z}_j \in \mathcal{M}_j$, the policy $\mathbf{Z}_i^*$ is an $\mathcal{M}-$ dominating policy for all $N \geq N^*$. From Lemma (\ref{Lem:Dual_condtn1}) and Lemma (\ref{Lem:Dual_condtn2}), $N^* = \max\{N_1,N_2\}$ is independent of the channel distribution and power constraint $\pi_j, \overline{G}_i\;\forall\; 1 \leq j \leq N $ as long as the regularity conditions hold true. This ensures that $\mathbf{Z}_i^*$ is an $\mathcal{M}-$ dominating policy for all $1\leq i \leq N$ and all $N \geq N^*$.

\section{Properties of $\mathcal{M}_{i}$}
\label{propb}
\begin{lem}\label{lem:Best_response_set_Prop_1}
If $\overline{G}_{i} \geq (1-\pi_i(1))g_{max}$, then $\mathcal{M}_{i} = \{\mathbf{Z}^*_i\}$, where
\begin{equation}
Z^*_{i}(h,g)=\begin{cases}
\pi_i(h), & \text{if }g = g_{max}, ~h\neq 0\\
0, & \text{otherwise}.
\end{cases}\label{eq:Large_power_conns_IDSP_soln}
\end{equation}

\begin{proof}
When the power constraint ($\overline{G}_{i}$) is greater than the maximum allowed power for non-zero channel gains, the only best response policy is to transmit with the maximum power possible ($g_{max}$) for non-zero channel gains.
\end{proof}
\end{lem}

\begin{lem}\label{lem:Best_response_set_Prop_2}
If $\overline{G}_{i} < (1-\pi_i(0))g_{max}$, then for any policy $\mathbf{Z}_{i} \in \mathcal{M}_{i}$, $\overline{G_i(\textbf{Z}_i)}=\overline{G}_i.$ 
\begin{proof}
Let $\mathbf{Z}^*_{i} \in \mathcal{M}_{i}$.
When $\overline{G}_{i} < (1-\pi_i(0))g_{max}$, there exists at least one channel gain $h^{(k)}$ such that $c \triangleq \mathbb{E}[G_i|H_i=h^{(k)}]< g_{max}$. 
Now, consider a policy $\hat{\mathbf{Z}}_{i}$ such that 
\[
\hat{Z}_{i}(h^{(a)},g):=\begin{cases}
\pi_i(h^{(k)})\left(1-\frac{c}{g_{max}}\right) & \text{if }a=k,\; g = 0\\
\pi_i(h^{(k)})\left(\frac{c}{g_{max}}+\delta\right), & \text{if }a=k,\; g = g_{max}\\
{Z}_{i}^{*}(h^{(a)},g), & \text{otherwise},
\end{cases}
\]
where $\delta >0$ and $\pi_i(h^{(k)})\delta < \overline{G}_i- \overline{G}_i(\mathbf{Z}_i)$.
It can be seen that $\hat{\mathbf{Z}}_{i}\in\mathcal{Z}_i$ and $T_i(\hat{\mathbf{Z}}_i,\mathbf{Z}_{-i})-T_i(\mathbf{Z}^*_i,\mathbf{Z}_{-i})>0$.
This contradicts the fact that $\mathbf{Z}^*_{i} \in \mathcal{M}_{i}$; hence, $\overline{G_{i}(\mathbf{Z}_{i}^{*})}=\overline{G}_{i}$.
\end{proof}
\end{lem}

\begin{thm}
\label{lem:inf}
Let $\mathbf{Z}_i \in \mathcal{M}_i $, When the regularity conditions hold, we have  $\mu = \inf_{i \geq 1}\mathbb{E}[N(h_i,g_i)]  > 0$.
\end{thm}

\begin{proof}
Using Lemma (\ref{lem:Best_response_set_Prop_1}), we can bound 
{\scriptsize
\begin{equation}\label{ARb_int1}
\mu_i \geq (1-\pi_i(h^{(1)}))N(h^{2},g_{max})> N(h^{2},g_{max})\eta . 
\end{equation}
}
when $\overline{G}_{i} \geq (1-\pi_i(0))g_{max}$. When $\overline{G}_{i} < (1-\pi_i(0))g_{max}$ by Lemma (\ref{lem:Best_response_set_Prop_2}), $\overline{G_{i}(\mathbf{Z}_{i}^{*})}= \overline{G}_{i}> 0$. To obtain a lower bound on $\mathbb{E}[N(h_i,g_i)]$, we observe that
{\scriptsize
\begin{eqnarray}\label{Prob_bnd_one}\nonumber
&&\mathbb{E}[N(h_i,g_i)] = \sum_{k=2}^{K}\pi_i(h^{k})\int_{0}^{g_{max}}N(h^{(k)}g_i)dF_{i,k}(g_i).\\
&\geq&\mathbb{E}_Q[N(h^{(2)},G_i)] \triangleq \int_{0}^{g_{max}}N(h^{(2)}g_i)\left(\sum_{k=2}^{K}\pi_i(h^{k})dF_{i,k}(g_i)\right), 
\end{eqnarray}
}
where we define the distribution $Q(g)$ on $[0,g_{max}]$ as 
{\scriptsize
\begin{equation}\label{Prob_bnd_two}
Q(g) \triangleq \sum_{k=1}^{K}\pi_i(h^{k})F_{i,k}(g_i),
\end{equation}
}
 with the distribution for $k=1$ satisfies $F_{i,1}(0) = 1$. Further we denote $\mathbb{E}_Q[\cdot]$ as expectation with respect to the distribution $Q$. Thus we have 
{\scriptsize
\begin{equation}\label{Prob_bnd_three}
\mathbb{E}_Q[N(h^{(2)},G_i)] \triangleq \int_{0}^{g_{max}}N(h^{(2)}g_i)\left(\sum_{k=2}^{K}\pi_i(h^{k})dF_{i,k}(g_i)\right)
\end{equation}
}
and
{\scriptsize
\begin{equation}\label{Prob_bnd_four}
\mathbb{E}_Q[G_i] \triangleq \int_{0}^{g_{max}}g\left(\sum_{k=2}^{K}\pi_i(h^{k})dF_{i,k}(g)\right) = \overline{G}_i(\mathbf{Z}_i) = \overline{G}_i.
\end{equation}
}
Further we have $\mathbb{E}_Q[G^2_i] \leq g^2(A)$. Now using Paley-Zygmund Inequality \cite{apostol1974mathematical}, we have 
{\scriptsize
\begin{equation}\label{Prob_bnd_five}
Q\left(G_i\geq \frac{\mathbb{E}_Q[G_i]}{2}\right)\geq \frac{(\mathbb{E}_Q[G_i]^2}{4\mathbb{E}_Q[G_i^2]}\geq \frac{\overline{G}_i^2}{4g^2(A)}.
\end{equation}
}
Now from (\ref{Prob_bnd_one}) and(\ref{Prob_bnd_five}), we have  
{\scriptsize
\begin{eqnarray}\label{Prob_bnd_six}\nonumber
&\mathbb{E}[N(h_i,g_i)]& \geq \mathbb{E}_Q[N(h^{(2)},G_i)] \geq \mathbb{E}_Q\left[N(h^{(2)},G_i)\mathbf{1}_{\left\{G_i \geq \frac{\mathbb{E}_Q[G_i]}{2}\right\}}\right] \\\nonumber
&\geq & N\left(h^{(1)},\frac{\mathbb{E}_Q[G_i]}{2}\right)Q\left(G_i\geq \frac{\mathbb{E}_Q[G_i]}{2}\right)\\
 &\geq & N\left(h^{(1)},\frac{\overline{G}_i}{2}\right)\frac{\overline{G}_i^2}{4g^2(A)} \geq N\left(h^{(1)},\frac{\overline{G}}{2}\right)\frac{\overline{G}^2}{4g^2(A)}, 
\end{eqnarray}
}
where 
{\scriptsize
\begin{equation}
\mathbf{1}_{\left\{G_i \geq \frac{\mathbb{E}_Q[G_i]}{2}\right\}}
 = \begin{cases}
1 & \text{when } G_i \geq \frac{\mathbb{E}_Q[G_i]}{2}\\
0 & \text{otherwise.}
\end{cases}
\end{equation}
}
Thus we have from (\ref{ARb_int1}) and (\ref{Prob_bnd_six})
{\scriptsize 
\begin{equation}
\mu_i = \mathbb{E}[N(h_i,g_i)]\geq \min\left\{N(h^{2},g_{max})\eta, N\left(h^{(1)},\frac{\overline{G}}{2}\right)\frac{\overline{G}^2}{4g^2(A)} \right\}
\end{equation}
}
, hence $\mu = \inf_i \mu_i >0$ under the regularity conditions.

\end{proof}

\begin{thm}
\label{lem:Chernoff_bds}
When $\mathbf{Z}_{j}\in\mathcal{{M}}_{j}$ for $j=1,\cdots,N$, we have 
{\scriptsize
\begin{equation}
\mathbb{E}\left[\left(\sum_{j=1}^{N}N(h_j,g_j)+N_0\right)^{-k}\right] = \Theta\Bigg(\frac{1}{N^{k}}\Bigg).\label{eq:Chernoff_bds_1}
\end{equation}
}
Further, the order constants only depend upon $\mu$ and $N(h^{(K)},g_{max})$. 
\end{thm}

\begin{proof}
Let $Z=\sum_{j=1}^{N}N(h_j,g_j)$ and $N_0 > 0$. 
As $N(h_j,g_j)\leq N(h^{(K)},g_{max})$, we have
{\small
\begin{equation*}
\mathbb{\mathbb{E}}\left[\left(Z+N_0\right)^{-k}\right]\geq \left( N \cdot N(h^{(K)},g_{max}) + N_0\right)^{-k}= \Omega\left(\frac{1}{N^{k}}\right).
\end{equation*}
}
{\scriptsize
\begin{eqnarray*}
\mathbb{\mathbb{E}}(Z+N_{0})^{-k}\hspace{-3mm}&=&
\int_{-\infty}^{\frac{N\mu}{2}}(Z+1)^{-k}\Pr(Z)dZ+\int_{\frac{N\mu}{2}}^{\infty}(Z+N_0)^{-k}\Pr(Z)dZ,\\
&\leq&  \Pr\left(Z/N<\mu/2\right) + \frac{2^{k}}{\left(N\mu\right)^{k}},\\
&\overset{(a)}{\leq}& 2\exp\left(\frac{-N\mu^{2}}{2(N(h^{(K)},g_{max}))^2}\right)+ \frac{2^{k}}{\left(N\mu\right)^{k}}= O\left(\frac{1}{N^{k}}\right),
\end{eqnarray*}
}
where $(a)$ follows from Hoeffding's inequality and, by Theorem \ref{lem:inf}, $\mathbb{{E}}(N(h_j,g_j))\geq\mu>0$.
\end{proof}

\section{Analysis at large interference}
\label{Max_E_Y_j}
  In this section, under the regularity constraints and under the assumption of convexity, we prove Lemma (\ref{Dual_condtn_1}) and Lemma (\ref{Dual_condtn_2}). We first establish some preliminary results required.

\begin{lem}\label{Lem:g1_and_g2_exist}
Let $N(h,g)$ be strictly convex in $g \geq 0$, for all $h>0$. Then the following conditions hold true.
\begin{enumerate}
\item $N(h,g)>0$ for all $g> 0$, for all $h>0$. 
\item There exist $0< g_1 \leq g_2 < g_{max}$ and an $\epsilon >0,\; L > 0$ such that for all $h^{(k)},2 \leq k\leq K$, and any $g_2 < g < g_{max}$,
{\scriptsize
\begin{equation}\label{g_2_exist}
\left(\frac{N(h,g_{max})}{g_{max}}+\epsilon\right) \leq \frac{N(h^{(k)},g_{max}) - N(h^{(k)},g)}{g_{max}-g} \leq (L+\epsilon),
\end{equation}
}
and for $h=h^{(\tau)}$ and any $0 < g < g_1$,
{\scriptsize
\begin{equation}\label{g_1_exist}
0 \leq \frac{N(h^{(\tau)},g)}{g}  \leq \left(\frac{N(h^{(\tau)},g_{max})}{g_{max}}-\epsilon\right).
\end{equation}
}
\item  For any $ h > 0$ and any  $\delta > 0$, there exist a $\epsilon > 0$ such that, 
{\scriptsize
\begin{equation}\label{End_now}
\inf_{\delta\leq g\leq g_{max} -\delta}\frac{N(h,g_{max})g}{g_{max}} - N(h,g) > \epsilon.
\end{equation}
}
\end{enumerate}
\end{lem}
\begin{proof}

Fix a $h \in \mathcal{H}/{0}$. By convexity for any $0 \leq g \leq g_{max}$ we have $N(h,g) < \frac{gN(h,g_{max})}{g_{max}}$. Further by  the monotonically increasing behaviour of $N(h,g)$, we have $N(h,g) > 0$ $\forall g \in (0,g_{max}]$. If not, then there exist a $g'>0$, such that $N(h,g) = 0, \forall\; g \in [0,g']$, but by strict convexity, we get $N(h,g'/2)< N(h,g')/2 < 0$. As this contradicts the monotonicity of $N(h,g)$, we must have $g'=0$, and thus $N(h,g) > 0$ $\forall g \in (0,g_{max}]$. This establishes part (1).

To establish part(2), we first establish that for all $h>0$, we have  
{\scriptsize
\begin{equation}\label{Prop_at_0}
0 \leq \lim_{g \rightarrow 0} \frac{N(h,g)}{g} < \frac{N(h,g_{max})}{g_{max}}.
\end{equation}
}
and for all $h>0$ and some $L>0$,
{\scriptsize
\begin{equation}\label{Prop_at_gmax}
\frac{N(h,g_{max})}{g_{max}} < \lim_{g \rightarrow 0} \frac{N(h,g_{max})-N(h,g)}{g_{max}-g} \leq L
\end{equation}
}
  
We first note that by  strict convexity \cite{boyd2004convex}, we have for any  real valued convex function $f(g), g \in \mathbb{R}$ and points $a< b < c$,
 \begin{equation}\label{Convex_3}
 \frac{f(b)-f(a)}{b-a} < \frac{f(c)-f(a)}{c-a}< \frac{f(c)-f(b)}{c-b}
 \end{equation}.
 By choosing a small enough $\delta > 0$, $f(g) = N(h,g)$, $a =0$, $b =  \delta$, $c= g_{max}$, we have from (\ref{Convex_3}) 
 \begin{equation}\label{Convex_4}
 \frac{N(h,\delta)}{\delta} < \frac{N(h,g_{max})}{g_{max}}.
 \end{equation} 
 and by choosing $a=0$, $b= g \in (0,\delta)$ and $c = \delta$, we have from (\ref{Convex_3})
 \begin{equation}\label{Convex_5}
 \frac{N(h,g)}{g} < \frac{N(h,\delta)}{\delta}.
 \end{equation} 
 From (\ref{Convex_4}) and (\ref{Convex_5}), we get (\ref{Prop_at_0}).
 Now let $L(h) = \lim_{g \rightarrow g_{max}}\frac{N(h,g_{max})-N(h,g)}{g_{max}-g}$ and $L = \max_{h>0}L(h)$. Then by a proof similar to that for (\ref{Prop_at_0}), we get (\ref{Prop_at_gmax})
Using the condition (\ref{Prop_at_gmax}), for some $\epsilon'_k > 0, 1\leq k\leq K$, we can find $0< g'_k < g_{max}$ such that for all $2 \leq k \leq K$ and $g'_k < g < g_{max}$
{\scriptsize
\begin{equation}
\left(\frac{N(h,g_{max})}{g_{max}}+\epsilon'_k\right) \leq \frac{N(h^{(k)},g_{max}) - N(h^{(k)},g_{max})}{g_{max}-g} \leq (L+\epsilon'_1),
\end{equation}
}
We define $\epsilon_2 = \min\limits_{1 \leq k\leq K}\epsilon'_k >0$ and $g_2 = \max\limits_{2 \leq k\leq K}g'_k$. Further using condition (\ref{Prop_at_0}), for some $\epsilon_1 > 0$, there exist a $0< g' < g_{max}$ such that for all $0 < g < g'$
\begin{equation}
0 \leq N(h^{(\tau)},g)  \leq \left(\frac{N(h^{(\tau)},g_{max})}{g_{max}}-\epsilon_1\right)g.
\end{equation}
We set $g_1 = \min\{g_2,g'\}$ and $\epsilon = \min\{\epsilon_1,\epsilon_2\} > 0 $ to get the desired results.

Now we prove part $(3)$. Let $\delta\leq g\leq g_{max} -\delta$ for some $\delta > 0$  and let $\alpha = \frac{g_{max}-\delta -g}{g_{max}-2\delta}$, then by strict convexity, we have for all $\delta\leq g\leq g_{max} -\delta$,
 {\scriptsize
 \begin{equation}\label{Convex_1}
 N(h,g) = N(h,\alpha \delta + (1-\alpha)(g_{max}-\delta))\leq \alpha N(h,\delta) + (1-\alpha)N(h,g_{max}-\delta).
 \end{equation}
 } 
  Let $\epsilon =$
 {\scriptsize
 \begin{equation}\label{Convex_2}
  min\left(\frac{N(h,g_{max})(g_{max}-\delta)}{g_{max}}-N(h,g_{max}-\delta),\frac{N(h,g_{max})(\delta)}{g_{max}}-N(h,\delta)\right). 
 \end{equation} 
 }
 Note that $\epsilon > 0$ by strict convexity. 
 Now  using (\ref{Convex_1}) and (\ref{Convex_2}),  for any $\delta\leq g\leq g_{max} -\delta$  we have $N(h,g) + \epsilon \leq \frac{N(h,g_{max})g}{g_{max}}$.
\end{proof}

\begin{lem}\label{Lem:Dual_condtn1}
Let $\mathbf{Z}_j \in \mathcal{M}_j$, $\forall \; 1\leq j\leq N$. Then under the regularity conditions, there exist a natural number $N_1$ independent of transmitter statistics and power constraints such that for $N \geq N_1$, 
{\scriptsize
\begin{eqnarray}\label{Dual_condtn_3_rep}
 0 &\geq & \triangleq \mathbb{E}_{-i}\left[\log\left(1 +\frac{N(h^{(k)},g)}{1 + \sum_{j \neq i} N(h_j,g_j)}\right)\right]\\\nonumber
&-&\frac{g}{g_{max}}\mathbb{E}_{-i}\left[\log\left(1 +\frac{N(h^{(\tau)},g_{max})}{1 + \sum_{j \neq i} N(h_j,g_j)}\right)\right], \\\nonumber
&\forall& g \in [0,g_{max}], 1 \leq k \leq \tau
\end{eqnarray}
}
\end{lem}
\begin{proof}
Note that as $N(h^{(k)},g) \leq N(h^{(\tau)},g) $, it suffices to prove (\ref{Dual_condtn_3_rep}) only for the case $k = \tau$.
We prove the result in three cases using the results of lemma (\ref{Lem:g1_and_g2_exist})
\begin{enumerate}
\item  $(g>g_2)$: Here we show there exist a integer $N_u$ such that $\forall N \geq N_u$ and $g_2 \leq g < g_{max}$,the condition (\ref{Dual_condtn_3_rep}) holds true. First by adding and subtracting the term $\mathbb{E}_{-i}\left[\log\left(1 +\frac{N(h^{(\tau)},g_{max})}{1 + \sum_{j \neq i} N(h_j,g_j)}\right)\right]$ to the term $T_1$ in (\ref{Dual_condtn_3_rep}), and rearranging, we get
{\scriptsize
\begin{eqnarray}\label{part1_1}
&T_1=-& \mathbb{E}_{-i}\left[\log\left(1 +\frac{N(h^{(\tau)},g_{max})-N(h^{(\tau)},g)}{1 + \sum_{j \neq i} N(h_j,g_j) + N(h^{(\tau)},g) }\right)\right]\\\nonumber
&+& \Bigg(\frac{(g_{max}-g)N(h^{(\tau)},g_{max})}{g_{max}}\\\nonumber
&\cdot&\mathbb{E}_{-i}\left[\log\left(1 +\frac{N(h^{(\tau)},g_{max})}{1 + \sum_{j \neq i} N(h_j,g_j)}\right)\right]\Bigg).
\end{eqnarray}
} 
Using (\ref{log_inequailities}) in (\ref{part1_1}), we get
{\scriptsize
 \begin{eqnarray}\label{part1_2}
 &T_1& \leq -\mathbb{E}_{-i}\left[ \frac{N(h^{(\tau)},g_{max})-N(h^{(\tau)},g)}{1 + \sum_{j \neq i} N(h_j,g_j) + N(h^{(\tau)},g) }\right]\\\nonumber
&+& \mathbb{E}_{-i}\left[ \frac{(N(h^{(\tau)},g_{max})-N(h^{(\tau)},g))^2}{2(1 + \sum_{j \neq i} N(h_j,g_j) + N(h^{(\tau)},g))^2 }\right]\\\nonumber
&+& \frac{(g_{max}-g)}{g_{max}}\mathbb{E}_{-i}\left[ +\frac{N(h^{(\tau)},g_{max})}{1 + \sum_{j \neq i} N(h_j,g_j)}\right].
\end{eqnarray}
} 
Using Lemma (\ref{Lem:g1_and_g2_exist}) in (\ref{part1_2}), we have for some $\epsilon>0$ and $0<g_2\geq g < g_{max}$, we get after rearranging terms 
{\scriptsize
 \begin{eqnarray}\label{part1_3}
 &T_1& \leq -\mathbb{E}_{-i}\left[ \frac{\epsilon(g_{max}-g)}{1 + \sum_{j \neq i} N(h_j,g_j) + N(h^{(\tau)},g) }\right]\\\nonumber
&+& \mathbb{E}_{-i}\left[ \frac{\left((L+\epsilon)(g_{max}-g)\right)^2}{2(1 + \sum_{j \neq i} N(h_j,g_j) + N(h^{(\tau)},g))^2 }\right] + \Bigg(\frac{(g_{max}-g)}{g_{max}}\cdot\\\nonumber
&\mathbb{E}_{-i} &\left[ \frac{N(h^{(\tau)},g_{max})N(h^{(\tau)},g)}{\left(1 + \sum_{j \neq i} N(h_j,g_j)\right)\left(1 + \sum_{j \neq i} N(h_j,g_j) + N(h^{(\tau)},g)\right)}\right]\Bigg).
\end{eqnarray}
}
Bounding $0 \leq N(h^{(\tau)},g) \leq N(h^{k},g_{max})$, we get from (\ref{part1_3}),
{\scriptsize
 \begin{eqnarray}\label{part1_4}
 &T_1& \leq -\mathbb{E}_{-i}\left[ \frac{\epsilon(g_{max}-g)}{1 + \sum_{j \neq i} N(h_j,g_j) + N(h^{(K)},g_{max}) }\right]\\\nonumber
&+& \mathbb{E}_{-i}\left[ \frac{\left(\left((L+\epsilon)g_{max}\right)^2 + N^2(h^{(K)},g_{max})\right)(g_{max}-g)}{g_{max}(1 + \sum_{j \neq i} N(h_j,g_j))^2 }\right] \\\nonumber
\end{eqnarray}
}
Using Theorem (\ref{lem:Chernoff_bds}) we get from (\ref{part1_4}),
\begin{equation}
T_1 \leq \left(\frac{-\epsilon c_1}{N} + \frac{c_2B_1}{N^2}\right)(g_{max}-g)
\end{equation} 
where the constants $c_1>0$ and $c_2>0$ depend only on $N(h^{(K)},g_{max})$ and $\mu$ which are fixed quantities, and $B_1 = \frac{\left((L+\epsilon)g_{max}\right)^2 + N^2(h^{(K)},g_{max})}{g_{max}}$.
Hence for all $g_2 \leq g \leq g_{max}$, there exist a $N_u$ such that for all $N\geq N_u$, $T_1 \leq 0$. 
\item  $(g<g_1)$: Here we show there exist a integer $N_l$ such that $\forall N \geq N_l$ and $0 \leq g < g_1$,the condition (\ref{Dual_condtn_3_rep}) holds true.
Using (\ref{log_inequailities}), we can bound the term $T_1$ in (\ref{Dual_condtn_3_rep}) as 
{\scriptsize
\begin{eqnarray}\label{Part2_1}
& T_1\leq &\left(N(h^{(\tau)},g)
-\frac{gN(h^{(\tau)},g_{max})}{g_{max}}\right)\mathbb{E}\left[\frac{1}{1 + \sum_{j \neq i} N(h_j,g_j)}\right], \\\nonumber
& + & \frac{g}{g_{max}}\mathbb{E}_{-i}\left[\frac{N^2(h^{(\tau)},g_{max})}{2\left(1 + \sum_{j \neq i} N(h_j,g_j)\right)}\right]
\end{eqnarray}
}
Using lemma (\ref{Lem:g1_and_g2_exist}) and bounding $N^2(h^{(\tau)},g_{max}) \leq N^2(h^{(K)},g_{max})$, we get in (\ref{Part2_1})
{\scriptsize
\begin{eqnarray}\label{Part2_2}
 T_1\leq \mathbb{E}_{-i}\left[\frac{-\epsilon g}{1 + \sum_{j \neq i} N(h_j,g_j)}\right]
+ \frac{g}{g_{max}}\mathbb{E}_{-i}\left[\frac{N^2(h^{(K)},g_{max})}{2\left(1 + \sum_{j \neq i} N(h_j,g_j)\right)}\right]
\end{eqnarray}
}
Using Theorem (\ref{lem:Chernoff_bds}) we get from (\ref{Part2_2}),
\begin{equation}
T_1 \leq \left(\frac{-\epsilon c_1}{N} + \frac{c_2B_2}{N^2}\right)g
\end{equation} 
where the constants $c_1$ and $c_2$ depend only on $N(h^{(K)},g_{max})$ and $\mu$ which are fixed quantities, and $B_2 = \frac{  N^2(h^{(K)},g_{max})}{2g_{max}}$.
Hence for all $0 \leq g \leq g_1$, there exist a $N_l$ such that for all $N\geq N_l$, $T_1 \leq 0$.

\item $(g_1 \leq g \leq g_2)$:Here we show there exist a integer $N_m$ such that $\forall N \geq N_m$ and $g_1 \leq g < g_2$,the condition (\ref{Dual_condtn_3_rep}) holds true.
Let $\delta = \min\{g_1,g_{max}-g_2\} $, then using condition (4) of definition there exist an $\epsilon_1>0$, such that
{\scriptsize 
\begin{equation}\label{part_3_1}
\left(N(h^{(\tau)},g)
-\frac{gN(h^{(\tau)},g_{max})}{g_{max}}\right) < -\epsilon_1.
\end{equation}
}
Substituting (\ref{part_3_1}) in (\ref{Part2_1}) and using that $g \leq g_{max}$ we get
{\scriptsize
\begin{eqnarray}\label{Part3_2}
 T_1\leq \mathbb{E}_{-i}\left[\frac{-\epsilon_1 }{1 + \sum_{j \neq i} N(h_j,g_j)}\right]
+ \mathbb{E}_{-i}\left[\frac{N^2(h^{(K)},g_{max})}{2\left(1 + \sum_{j \neq i} N(h_j,g_j)\right)}\right].
\end{eqnarray}
}
Similar to part $(2)$, there exist a $N_m$ such that for all $N\geq N_m$, $T_1 \leq 0$. Now by choosing $N_1 = \max\{N_l,N_u,N_m\}$, we ensure (\ref{Dual_condtn_3_rep}) holds for all $N\geq N_1$. Further as the integers $N_l,N_u,N_m$ only depend on $N(h^{K},g_{max}), g_{max}$ and $\mu$ and fixed constants $g_1,g_2,\epsilon,\epsilon_1$, the number $N_1$ is  independent of user statistics and power constraints.

\end{enumerate}

\end{proof}

\begin{lem}\label{Lem:Dual_condtn2}
Let $\mathbf{Z}_j \in \mathcal{M}_j$, $\forall 1\leq j\leq N$. Further let the regularity conditions hold true, then there exist a natural number $N_2$ independent of user statistics and power constraint such that for $N \geq N_2$, 
{\scriptsize
\begin{eqnarray}\label{Dual_condtn_4_rep}
&&\mathbb{E}_{-i}\left[\log\left(1 +\frac{N(h^{(k)},g_{max})}{1 + \sum_{j \neq i} N(h_j,g_j)}\right)\right] \\\nonumber
&-& \mathbb{E}_{-i}\left[\log\left(1 +\frac{N(h^{(\tau)},g_{max})}{1 + \sum_{j \neq i} N(h_j,g_j)}\right)\right]\\\nonumber
 &\geq& \mathbb{E}_{-i}\left[\log\left(1 +\frac{N(h^{(k)},g)}{1 + \sum_{j \neq i} N(h_j,g_j)}\right)\right]\\\nonumber
&-&\frac{g}{g_{max}}\mathbb{E}_{-i}\left[\log\left(1 +\frac{N(h^{(\tau)},g_{max})}{1 + \sum_{j \neq i} N(h_j,g_j)}\right)\right], \\\nonumber
&\forall& g \in [0,g_{max}], \tau + 1 \leq k \leq K.
\end{eqnarray}
}
\end{lem}
\begin{proof}
We prove the result in two cases using the results of lemma (\ref{Lem:g1_and_g2_exist})
\begin{enumerate}
\item $(g > g_2)$: First note that by rearranging the terms showing (\ref{Dual_condtn_4_rep}) is true is equivalent to showing 
{\scriptsize
\begin{eqnarray}\label{part4_1}
&0& \leq T_2 \triangleq \mathbb{E}_{-i}\left[\log\left(1 +\frac{N(h^{(k)},g_{max})-N(h^{(k)},g)}{1 + \sum_{j \neq i} N(h_j,g_j) + N(h^{(k)},g)}\right)\right]\\\nonumber
&-& \frac{(g_{max}-g)}{g_{max}}\mathbb{E}_{-i}\left[\log\left(1 +\frac{N(h^{(\tau)},g_{max})}{1 + \sum_{j \neq i} N(h_j,g_j)}\right)\right]
\end{eqnarray}
}
Applying (\ref{log_inequailities}) to (\ref{part4_1}), we get
{\scriptsize
\begin{eqnarray}\label{part4_2}
 &T_2& \geq \mathbb{E}_{-i}\left[\frac{N(h^{(k)},g_{max})-N(h^{(k)},g)}{1 + \sum_{j \neq i} N(h_j,g_j) + N(h^{(k)},g)}\right]\\\nonumber
 &-& \mathbb{E}_{-i}\left[\left(\frac{N(h^{(k)},g_{max})-N(h^{(k)},g)}{1 + \sum_{j \neq i} N(h_j,g_j) + N(h^{(k)},g)}\right)^2\right]\\\nonumber
&-& \frac{(g_{max}-g)}{g_{max}}\mathbb{E}_{-i}\left[\frac{N(h^{(\tau)},g_{max})}{1 + \sum_{j \neq i} N(h_j,g_j)}\right]
\end{eqnarray}
}
Using Lemma (\ref{Lem:g1_and_g2_exist}) in (\ref{part4_2}), we get for all $g_2 \leq g \leq g_{max}$, 
{\scriptsize
\begin{eqnarray}\label{part4_3}
 &T_2& \geq \mathbb{E}_{-i}\left[\frac{\epsilon(g_{max}-g)}{1 + \sum_{j \neq i} N(h_j,g_j) + N(h^{(k)},g)}\right]\\\nonumber
 &+& \mathbb{E}_{-i}\left[\frac{N(h^{(k)},g_{max})(g_{max}-g)}{g_{max}\left(1 + \sum_{j \neq i} N(h_j,g_j) + N(h^{(k)},g)\right)}\right]\\\nonumber 
 &-& \mathbb{E}_{-i}\left[\left(\frac{(L+\epsilon)(g_{max}-g)}{1 + \sum_{j \neq i} N(h_j,g_j) + N(h^{(k)},g)}\right)^2\right]\\\nonumber
&-& \frac{(g_{max}-g)}{g_{max}}\mathbb{E}_{-i}\left[\frac{N(h^{(\tau)},g_{max})}{1 + \sum_{j \neq i} N(h_j,g_j)}\right]
\end{eqnarray} 
}
Bounding $N(h^{(\tau)},g_{max}) < N(h^{(k)},g_{max}) $, for $k > \tau$ and $0 \leq N(h^{(k)},g) < N(h^{(K)},g_{max})$ , we get in (\ref{part4_3}), we get 
{\scriptsize
\begin{eqnarray}\label{part4_4}
 &T_2& \geq \mathbb{E}_{-i}\left[\frac{\epsilon(g_{max}-g)}{1 + \sum_{j \neq i} N(h_j,g_j) + N(h^{(K)},g_{max})}\right]\\\nonumber
 &-& \mathbb{E}_{-i}\left[\frac{N(h^{(\tau)},g_{max})N(h^{(k)},g)(g_{max}-g)}{g_{max}\left(1 + \sum_{j \neq i} N(h_j,g_j)  \right)}\right]\\\nonumber 
 &-& \mathbb{E}_{-i}\left[\left(\frac{(L+\epsilon)(g_{max}-g)}{1 + \sum_{j \neq i} N(h_j,g_j) }\right)^2\right]\\\nonumber
\end{eqnarray} 
}
 Further bounding using $0 \leq N(h^{(k)},g) < N(h^{(K)},g_{max})$, we get,
{\scriptsize
\begin{eqnarray}\label{part4_5}
 &T_2& \geq \mathbb{E}_{-i}\left[\frac{\epsilon(g_{max}-g)}{1 + \sum_{j \neq i} N(h_j,g_j) + N(h^{(K)},g_{max})}\right]\\\nonumber
 &-& \mathbb{E}_{-i}\left[\frac{\left(((L+\epsilon)g_{max})^2)+ N^2(h^{(K)},g_{max}\right)(g_{max}-g)}{g_{max}\left(1 + \sum_{j \neq i} N(h_j,g_j)\right)^2 }\right]\\\nonumber
\end{eqnarray} 
}
Using Theorem (\ref{lem:Chernoff_bds}) we get from (\ref{part1_4}),
{\scriptsize
\begin{equation}
T_2 \geq \left(\frac{\epsilon c_1}{N} - \frac{c_2B_1}{N^2}\right)(g_{max}-g)
\end{equation} 
}
where the constants $c_1>0$ and $c_2>0$ depend only on $N(h^{(K)},g_{max})$ and $\mu$ which are fixed quantities, and $B_1 = \frac{\left((L+\epsilon)g_{max}\right)^2 + N^2(h^{(K)},g_{max})}{g_{max}}$.
Hence for all $g_2 \leq g \leq g_{max}$, and for all $k > \tau$, there exist a $N'_u$ such that for all $N\geq N'_u$, $T_2 \geq 0$. 
\item $(0\leq g\leq g_2)$: We first note that (\ref{Dual_condtn_4_rep}) is equivalent to showing
$T_3 - T_4 \geq 0$, where
{\scriptsize
\begin{eqnarray}\label{T_3_defn}
&T_3& \triangleq \mathbb{E}_{-i}\left[\log\left(1 +\frac{N(h^{(k)},g_{max})}{1 + \sum_{j \neq i} N(h_j,g_j)}\right)\right] \\\nonumber
&-& \mathbb{E}_{-i}\left[\log\left(1 +\frac{N(h^{(\tau)},g_{max})}{1 + \sum_{j \neq i} N(h_j,g_j)}\right)\right]\\\nonumber
\end{eqnarray}
}
and
{\scriptsize
\begin{eqnarray}\label{T_4_defn}
 &T_4& \triangleq \mathbb{E}_{-i}\left[\log\left(1 +\frac{N(h^{(k)},g)}{1 + \sum_{j \neq i} N(h_j,g_j)}\right)\right]\\\nonumber
&-&\frac{g}{g_{max}}\mathbb{E}_{-i}\left[\log\left(1 +\frac{N(h^{(\tau)},g_{max})}{1 + \sum_{j \neq i} N(h_j,g_j)}\right)\right], \\\nonumber
\end{eqnarray}
}
We have from (\ref{log_inequailities}) and $N(h^{(k)},g_{max}) \leq N(h^{(K)},g_{max})$,
{\scriptsize
\begin{eqnarray}\label{T_3_defn_1}
&T_3& \geq \mathbb{E}_{-i}\left[\frac{N(h^{(k)},g_{max})- N(h^{(\tau)},g_{max})}{1 + \sum_{j \neq i} N(h_j,g_j)}\right] \\\nonumber
&-& \mathbb{E}_{-i}\left[\left(\frac{N(h^{(K)},g_{max})}{1 + \sum_{j \neq i} N(h_j,g_j)}\right)^2\right]. \\\nonumber
\end{eqnarray}
}
Similarly We have from (\ref{log_inequailities}) and $N(h^{(k)},g_{max}) \leq N(h^{(K)},g_{max})$ in (\ref{T_4_defn}),
{\scriptsize
\begin{eqnarray}\label{T_4_defn_2}
 &T_4& \leq \mathbb{E}_{-i}\left[\frac{N(h^{(k)},g)}{1 + \sum_{j \neq i} N(h_j,g_j)}\right]\\\nonumber
&-&\frac{g}{g_{max}}\mathbb{E}_{-i}\left[\frac{N(h^{(\tau)},g_{max})}{1 + \sum_{j \neq i} N(h_j,g_j)})\right], \\\nonumber
&+&\frac{g}{g_{max}}\mathbb{E}_{-i}\left[\left(\frac{N(h^{(\tau)},g_{max})}{1 + \sum_{j \neq i} N(h_j,g_j)})\right)^2\right], \\\nonumber
\end{eqnarray}
}
Using the strict convexity of $N(h,g)$, $g \leq g_{max}$ and $N(h^{(\tau)},g_{max}) \leq N(h^{(K)},g_{max})$ we get
{\scriptsize
\begin{eqnarray}\label{T_4_defn_3}
 &T_4& \leq \frac{g}{g_{max}}\mathbb{E}_{-i}\left[\frac{N(h^{(k)},g_{max})-N(h^{(\tau)},g_{max})}{1+ \sum_{j \neq i} N(h_j,g_j)}\right]\\\nonumber
&+&\mathbb{E}_{-i}\left[\left(\frac{N(h^{(K)},g_{max})}{1 + \sum_{j \neq i} N(h_j,g_j)})\right)^2\right], \\\nonumber
\end{eqnarray}
}
 Now as $0 \leq g \leq g_2$, we get
{\scriptsize
\begin{eqnarray}\label{T_4_defn_4}
 &&T_3-T_4 \geq \frac{(g_{max}-g_2)}{g_{max}}\mathbb{E}_{-i}\left[\frac{N(h^{(k)},g_{max})-N(h^{(\tau)},g_{max})}{ \sum_{j \neq i} N(h_j,g_j)}\right]\\\nonumber
&+&2\mathbb{E}_{-i}\left[\left(\frac{N(h^{(K)},g_{max})}{1 + \sum_{j \neq i} N(h_j,g_j)})\right)^2\right], \\\nonumber
\end{eqnarray}
}
Further let $\epsilon_3 = \min\limits_{2 \leq k_1<k_2 \leq K}N(h^{(k_2)},g_{max})-N(h^{(k_1)},g_{max})$. As $\mathcal{H}$ is finite $\epsilon_3 >0$, Now we have in (\ref{T_4_defn_4})  
{\scriptsize
\begin{eqnarray}\label{T_4_defn_5}
 &&T_3-T_4 \geq \frac{(g_{max}-g_2)}{g_{max}}\mathbb{E}_{-i}\left[\frac{\epsilon_3}{ \sum_{j \neq i} N(h_j,g_j)}\right]\\\nonumber
&+&2\mathbb{E}_{-i}\left[\left(\frac{N(h^{(K)},g_{max})}{1 + \sum_{j \neq i} N(h_j,g_j)})\right)^2\right], \\\nonumber
\end{eqnarray}
}
Using Theorem (\ref{lem:Chernoff_bds}) we get from (\ref{T_4_defn_5}),
{\scriptsize
\begin{equation}
T_3-T_4 \geq \left(\frac{\epsilon_3 c_1}{N} - \frac{c_2B_1}{N^2}\right)
\end{equation} 
}
where the constants $c_1>0$ and $c_2>0$ depend only on $N(h^{(K)},g_{max})$ and $\mu$ which are fixed quantities, and $B_1 = \frac{N(h^{(K)},g_{max})g_{max}}{g_{max}-g_2}$.
Hence for all $0 \leq g \leq g_2$, and for all $k > \tau$, there exist a $N'_m$ such that for all $N\geq N'_m$, $T_3-T_4 \geq 0$. Now by choosing $N_2 = \max\{N'_u,N'_m\}$, we ensure (\ref{Dual_condtn_4_rep}) holds for all $N\geq N_2$. Further as the integers $N'_u,N'_m$ only depend on $N(h^{K},g_{max}), g_{max}$ and $\mu$ and fixed constants $g_2,\epsilon,\epsilon_3$, the number $N_2$ is  independent of user statistics and power constraints.

\end{enumerate}

\end{proof}

\end{appendices}
\end{document}